\documentclass[11pt]{article}
\usepackage{amsmath,epsfig,capt-of,ifthen,calc}

\begin{document}
\author{{\sc Dom\`enec Espriu}\footnote{espriu@ecm.ub.es} and 
{\sc Luca Tagliacozzo}\footnote{luca@ecm.ub.es}\\
Departament d'Estructura i Constituents de la Mat\`eria,
Universitat de Barcelona\\
Diagonal, 647, 08028 Barcelona, Spain}
\date{}
\title{\bf Compact lattice $U(1)$ and Seiberg-Witten duality:\\
a quantitative comparison }
\maketitle

\bigskip\bigskip
\begin{abstract}
It was conjectured some time ago that an effective description of the 
Coulomb-confinement transition in compact $U(1)$ lattice gauge field 
theory could 
be described by scalar QED obtained by 
soft breaking of the $N=2$ Seiberg-Witten model down to $N=0$ in the strong 
coupling region where monopoles are light. 
In two previous works this idea was presented at a qualitative level. 
In this work we analyze in detail the conjecture and obtain encouraging 
quantitative agreement with the numerical determination of the monopole 
mass and the dual photon mass 
in the vicinity of the Coulomb to confining phase transition. 
\end{abstract}

\vfill

\noindent
UB-ECM-PF-04/10

\noindent
April 2004

\newpage

The $U(1)$ lattice gauge theory in four dimensions has been long 
studied \cite{Creutz:1979zg}.
Among other interesting features it is the simplest theory posessing a
confining phase in four dimensions. This phase is separated from the  Coulomb
one by a phase transition that it is believed to be driven my monopole 
condensation. The order of such transition has been subject 
to debate for a long time.

The analysis of this phase transition are usually done with an extended
Wilson  action \cite{Bhanot:1981tn}
\begin{equation}
  S = - \sum_P ( \beta \cos \theta_P + \gamma \cos 2 \theta_P ) \hspace{2em}
  \theta_P \in [ 0, 2 \pi ). \label{lataction}
\end{equation}
It was generally believed that generic values of the coupling $\gamma$ 
lead to a Coulomb-confinement transition of first order,
with larger values of the coupling $\gamma$ yielding 
stronger transitions. 

With the possibility to study larger lattices and, accordingly,
to explore larger 
correlation lengths, from a numerical point of view, some
evidence for a second order nature of the transition were
found \cite{Jersak:1996mn,Cox:1997wd}, at least
for negative values of $\gamma$. The critical exponents and the 
spectrum were measured. The ligthest part of spectrum, according to the 
authors of \cite{Jersak:1996mn,Cox:1997wd} consists in a $0^+$ 
and a $1^+$ gaugeballs that appear to scale differently ($\nu=0.49(7)$
and $0.35(3)$ for $\gamma=-0.2$, respectively). Other states seem to be 
compatible with being multiple particle states. All states considered 
are created by local or quasilocal operators so they presumably 
carry no magnetic charge. These measurements have been repeated recently  
with high precision in the work \cite{Majumdar:2003xv}. 

A quite different set of measurements was performed by the Pisa group 
\cite{ DiGiacomo:1997sm}.
They explicitly construct operators that create and annihilate 
monopoles. They are non-local in the original (electrical) variables. These
authors also analyzed the spectrum and found the dependence of the
monopole ($0^+$) and the dual photon ($1^+$) on the gauge coupling in 
the confined phase, 
determining some effective critical indices. For the monopole they
find $\nu^\ast=0.29(2)$.  

Indeed it seems to be now universally accepted,
that the phase transition is 
actually of first order, albeit a very weak 
one {\cite{Campos:1998ds,Arnold:2002jk,Vettorazzo:2003fg}. 
This  indicates that the critical
exponents found in  \cite{Jersak:1996mn,Cox:1997wd}
have to be taken as effective and that very close to the phase
transition its first order nature should finally reveal.

Nevertheless, even if no true new fixed point is found, it is still very
interesting to understand analytically this transition as it is
driven by monopole condensation. This problem was addressed in two previous
works {\cite{Ambjorn:1997cq,Espriu:2003sa}}. 
These authors follow a continuum treatment (that will be used in the present 
work too) and this is justified a posteriori
because the transition exhibits a very large
correlation length, extending over many lattice spacings 
and therefore the cut-off effects are small, of ${\cal O}(a/\xi)$ if 
$\xi$ is the correlation length. The Pisa group has indeed checked that
some kind of universality exists by using scaling arguments.
Using a continuum language makes 
everything simpler. 

A priori it is not obvious how to write an effective field theory description
of the monopole field and its interactions.
It was proposed in  {\cite{Ambjorn:1997cq,Espriu:2003sa}}
to use as guiding principle the
formulation of $N = 2$ SYM by Seiberg and Witten {\cite{Seiberg:1994rs}}, 
suitably
adapted for this scenario; the idea was to use  a pattern of symmetry breaking
providing  at low energies the  spectrum  observed on the lattice. It was
found in {\cite{Espriu:2003sa}} that the effective theory previously 
formulated in {\cite{Ambjorn:1997cq}} can predict a phase diagram 
that qualitatively agrees with the one found in lattice simulations 
corresponding to a weakly first order transition.

In this work we aim to extend such agreement from the qualitative to the
quantitative level. We reproduce the plot of the mass of the monopole field
found in {\cite{DiGiacomo:1997sm}}  and check that in the confining phase its 
pseudo critical exponent $\nu$ is Gaussian. 
We also determine the mass of the dual photon state, that we actually use 
to match some continuum parameters to the lattice values. 

The key point (see \cite{Espriu:2003sa}) is to properly understand the 
scale dependence of the low-energy
effective theory. In its original formulation, derived  by Seiberg and
Witten, the effective action
provides the right description at a scale  $\vert a_D\vert$ 
in the dual variables ($a_D$ being the v.e.v. of the dual scalar 
field which is a
component of the original $N = 2$ vector multiplet). 
As already explained in {\cite{Espriu:2003sa}} at
the scale $\vert a_D\vert$  
the couplings freeze i.e. they stop running; this allows to extract  
the theory at $\mu \sim 0$
from the one at $\mu \sim \vert a_D\vert $. 

In order to describe a scenario with $U(1)$ gauge symmetry and no 
supersymmetry ($N=0$)  in {\cite{Ambjorn:1997cq}}  a two step
supersymmetry 
breaking was introduced using both the coupling  to a spurion
superfield 
that has a non-zero value for its $D$ 
component (a technique borrowed from {\cite{Alvarez-Gaume:1996gd}}) 
and to a $N = 1$ superfield.
The addition of the hard $N = 1$ breaking is absolutely crucial as 
it allows us to 
decouple the value of $a_D$
from the point in moduli space one has to choose to define the theory; this
allows  the monopole  to become light far from the original region in the 
(unperturbed) moduli space where  $a_D \sim 0$. 
In fact, the value for   $\vert a_D\vert $  for which the
monopole become massless turns out to be close to ${D}_0$, where $D_0$ is
the $D$ component of the spurion field that breaks $N = 2$ down to $N
= 0$. 
In spite of the hard nature of the $N=1$ breaking term, 
the dual version of the effective action 
can still be determined to some extent, and this is sufficient for our 
purposes. The analytic structure of the K\"ahler
prepotential $\mathcal{F}$ is unchanged.

Thus even if at the scale ($\mu \sim \vert a_D\vert )$  
the form of the effective potential
is uniquely determined from the Seiberg and Witten construction, 
at lower energies 
(the ones relevant for the  comparison with the
lattice results if we suppose the lattice scale at $\mu = 1$)  there are
corrections produced  by quantum effects due to the breaking of supersymmetry.
This is how  the Coleman-Weinberg
mechanism {\cite{Coleman:1973jx,Yamagishi:1981qq}} triggers a (weak) 
first order transition.

Before entering into the details of the calculation we review the
frameset where we work. This will also allow us to introduce the
notation we shall  use
in the subsequent. Starting from the $N = 2$ SYM {\cite{Seiberg:1994rs}} 
and breaking
all the supersymmetries down to $N = 0$ 
{\cite{Alvarez-Gaume:1996gd,Ambjorn:1997cq}} one obtains
an effective theory whose Lagrangian density is:

\begin{eqnarray}
  \mathcal{L} & = & - \frac{1}{4} F_{\mu \nu} F^{\mu \nu} + \frac{1}{2}
  \partial^{\mu} m \partial_{\mu} m + \frac{1}{2} \partial^{\mu} \theta
  \partial_{\mu} \theta \nonumber\\
  &  & + g_{_D} m \partial_{\mu} \theta A^{\mu} + g_{_D} \theta
  \partial_{\mu} mA^{\mu} + \frac{1}{2}g_D^2A^{\mu}A_{\mu} m^2\nonumber\\
  &  & - \frac{\lambda}{16}  m^4 - \frac{1}{2} M^2 m^2 +\mathcal{L}_{g.f.}
  \label{stlag}
\end{eqnarray}
The field
$A_{\mu}$ denotes the dual photon ($J^P = 1^+$) and $m, \theta$ are the polar 
components of the complex monopole field ($J^P = 0^+$).
The scalar potential is given by the two last terms in
(\ref{stlag}).
In the previous formulae only light fields have been retained. 
The full effective lagrangian contains in addition fields of masses
$\mathcal{O}( {D_0} )$, but we have not written these since we are
eventually interested in the energy range for $p^2$: $0 \le p^2 \ll b_{01}
D_0$, and in this range they  decouple (see {\cite{Espriu:2003sa}} for the full
spectrum). Thus the above lagrangian 
provides a valid description only below the  scale $\mu = {D_0}\simeq a_D$. 
The couplings $b_{ij}$
are expressed in terms of the original prepotential
\begin{equation}
  b_{ij} \equiv \frac{1}{4 \pi} \text{\textrm{Im}} \ \tau_{ij} =
  \frac{1}{4 \pi} \text{\textrm{Im}} \ \frac{\partial^2
  \mathcal{F}}{\partial a_D^i \partial a_D^j},
\end{equation}
the index `1' denoting the physical fields and `0', the spurion 
\cite{Alvarez-Gaume:1996gd}. 
An important parameter
is the monopole `mass' term. At the scale ${D_0}$
\begin{equation}
  M^2 \equiv -\frac{\alpha}{b_{11}} \label{mass}.
\end{equation}
Here
\begin{equation}
\alpha=-\left( 2|a_D|^2 + b_{01} D_0\right)
  \label{alpha}
\end{equation}

On the other hand the mass of the dual photon is given by the formula
familiar from the Higgs mechanism. At tree level, reading from 
(\ref{stlag}), $M_V=g_D\langle m\rangle$, where $\langle m\rangle$ is 
the expectation value derived from the tree level effective potential.  

Let us remark once more that the prepotential $\mathcal{F}$ is unchanged by
the $N = 1$ breaking {\cite{Ambjorn:1997cq}}. 
All the previous results hold at the 
supersymmetry breaking scale ${D_0} \sim
\vert a_D\vert $. At this scale supersymmetry implies
\begin{equation}
  g_D = \frac{1}{\sqrt{b_{11}}}, \hspace{2em} \lambda = \frac{2}{b_{11}}
  \label{relbar}
\end{equation}

Next we have to run the effective potential from the scale $\mu = {D_0}$
down to $\mu \sim 1$ to compare with the lattice results. This is done in 
perturbation theory  at the  one-loop approximation. The result is
\begin{eqnarray}
  &  & V = \frac{\lambda}{16} m^4 + \frac{M^2}{2} m^2 + \frac{1}{64 \pi^2}
  \left( \frac{\lambda m^2}{4} + M^2 \right)^2 \ln \left( \left( \frac{\lambda
  m^2}{4} + M^2 \right) \frac{1}{\mu^2} - \frac{3}{2} \right) + \nonumber\\
  &  & \frac{1}{64 \pi^2} \left( \hspace{0.25em} \frac{3 \lambda m^2}{4} +
  M^2 \right)^2 \ln \left( \left( \frac{3 \lambda m^2}{4} + M^2 \right)
  \frac{1}{\mu^2} - \frac{3}{2} \right) + \frac{3 m^4 g_D^4}{64 \pi^2} \ln
  \left( \frac{g_D^2m^2}{\mu^2} - \frac{5}{6} \right) . \label{renpot} 
\end{eqnarray}

We quote here the relevant beta functions \cite{Ahmed:1974sy}
\begin{equation}
  \beta_g = \frac{g_D^3}{48 \pi^2} \label{betag}
\end{equation}
\begin{equation}
  \beta_{\lambda} = \frac{5 \lambda^2}{16 \pi^2} - \frac{3 \lambda g_D^2}{4
  \pi^2} + \frac{3 g_D^4}{2 \pi^2} \label{betal}
\end{equation}
\begin{equation}
  \gamma_m = - \frac{3 g_D^2}{8 \pi^2} \label{betaz}
\end{equation}
\begin{equation}
  \gamma_M = - 2 + \frac{\lambda}{8 \pi^2} - \frac{3 g_D^2}{8 \pi^2}
  \label{betam}
\end{equation}
At first glance the lagrangian (\ref{stlag}) looks like standard
scalar electrodynamics but there are some important
remarks to be made. 
First of all, it is not even obvious
that scalar electrodynamics is the natural language to describe
the Coulomb-confinement transition. 
%(can monopoles be described by local variables? The consistency of
%the Seiberg-Witten construction tells us that the answer is
%positive). 
The  supersymmetric origin prescribes a
well defined relation between $g_D$ and $\lambda$ at the scale  $\mu =
{D_0}$ (\ref{relbar}) that can be extended to  any other scale by using
the  renormalization group equations via the above beta functions.
Furthermore  there is a  dependence of the  mass term $M^2$ on $g_D$ 
that is also  dictated 
by supersymmetry, and this turns out to be absolutely crucial to
provide a successful description (see below).

Likewise the dual photon mass algo gets renormalized when running down from
$\mu=D_0$ to $\mu=1$
\begin{equation}
M_V=g_D\langle m \rangle + \delta M_V
\end{equation}
Here $g_D$ is the (dual) gauge coupling at scale $\mu=1$, $\langle m\rangle$
is the v.e.v. derived from (\ref{renpot}) and $ \delta M_V$ is the one-loop 
correction. We do not present it here but  it has been included in
our analysis.

Let us have a closer look at   
the definition of $M^2$. From
(\ref{alpha}), using the results of 
\cite{Alvarez-Gaume:1996gd}, we have 
\begin{equation}
  M^2 =  \left( 2| a_D |^2 + \frac{\Re a_D}{4 \pi} D_0 \right)g_D^2 + \Im a D_0
\end{equation} On the
other hand as already explained in \cite{Espriu:2003sa} $|a_D|$ 
is the parameter 
that drives the phase transition whereas $a$ is related to the point 
on the moduli space of the original Seiberg-Witten construction.

As shown in detail in \cite{Ambjorn:1997cq} and in \cite{Espriu:2003sa},
the addition of the $N=1$ breaking allows the use of $|a_D|$ as a 
free parameter
triggering the monopole condensation. 
In fact, the value of $|a_D|$ is chosen so
as to reproduce the critical coupling in the lattice. The relation between 
$a$ and $a_D$ is not completely determined because it is modified by the 
hard breaking that the $N=1$ term brings about. The form of the relation 
is derived in \cite{Ambjorn:1997cq} using the charges under $U(1)_A\times
U(1)_R$ symmetry. 
The value of 
$\vert a_D\vert$ is assumed to be tuneable by adjusting the 
value of the dimensionful parameter in the $N=1$ hard breaking term (called $w$
in \cite{Espriu:2003sa}). Likewise the same mechanism should also
allow us to tune the value for $\Im a$ by using the
imaginary part of the same hard breaking parameter 
$w$ \cite{Ambjorn:1997cq,Espriu:2003sa} . In other words, we 
trade the two degrees of freedom of the complex parameter $w$
by $\vert a_D\vert$ and $\Im a$. It should be clear to the reader that
this is an assumption, though a very reasonable one.

Thus, for simplicity, we write the monopole mass parameter $M^2$ as
\begin{equation}
M^2 =  \alpha_0 + A g_D^2 \label{massbare}
\end{equation}
where $\alpha_0$ and $A$ (instead of $a_D$ and $D_0$) are adjustable  (real)
parameters in our approach to be matched to the lattice action. 
This should be valid up to higher orders in $g_D$ that do not really
affect the essence of the discussion.

Let us now make contact with the lattice results.
We shall identify 
the $g_D$ coupling with $\beta$ introduced in (\ref{lataction}): 
$g_D^2=\beta$. This is a reasonable assumption considering the
standard lattice-to-continuum relation $\frac{1}{g^2}=\beta$ in 
conjuction with the relation 
between direct and dual couplings $g_D=1/g$ and the fact that in the continuum 
the gauge coupling is associated to a conserved current.
The value of ${D_0}$, the supersymmetry breaking scale, as explained 
in detail in \cite{Espriu:2003sa},  can  be thought as the 
ultraviolet cut off for
the bare lagrangian (\ref{stlag}).

The continuum model hence depends on 4 parameters:
$g_D,D_0,\alpha_0$ and $A$. Our results depend very weakly 
on the value of $D_0$.
This is due to the fact that the lattice phase 
transition (and the continuum version of it) is weakly first order so 
in the range of energies we are considering  
(where the physical observables obey a pseudo scaling and a finite 
size analysis produces the typical  collapse of the data expected in a 
second order phase transition \cite{DiGiacomo:1997sm}) we expect a 
very weak dependence of our results on the UV cut-off $D_0$. 
This assumption has been checked by varying  the value of $D_0$. We 
must however ensure the validity of the 
one loop approximation for the beta functions.

Given these assumptions we are left with an effective theory depending on 
one parameter ($g_D$) that is directly related to the lattice model and 
two free parameters $\alpha_0,A$ that we should fix matching some 
physical observables between the lattice and the continuum results (actually
the lattice can be thought of a continuum theory with an infinite number
of higher dimensional operators too if we want to look at it in this way).

In the work of \cite{DiGiacomo:1997sm} two physical observables 
directly accessible in our model have been measured, namely
$M_{V}^{latt}(\beta)$  and $M^{latt}(\beta)$: the dual photon and the
monopole masses.
The notation emphasizes that they are lattice results. The measurement is 
reasonable for the dual photon mass, but
with enormous error bars for the monopole mass.
On the other hand,  
$M_{V}(g_D,\alpha_0,A),M(g_D,\alpha_0,A)$ 
refer to the curves we obtain in the continuum theory for the masses 
as a function of the free parameters.
In our model both of these curves can be easily calculated. Just to fix ideas,
the  parameters $g_D,\alpha_0$ and $M$ have to be understood at the
scale $D_0$. The renormalization group allows us to determine its value
at the relevant scale $\mu\sim 1$ as well as the rest of the parameters that
are related to them (such as $\lambda(\mu\sim 1)$)  

The monopole condensate for instance $ \langle m \rangle $ is obtained 
at the one loop level by minimizing the potential (\ref{renpot}).
The monopole mass $M^2(g_D,\alpha_0,A)$ is obtained as the 
second derivative of the potential (\ref{renpot}) with respect to $m$ 
field at its minimum $m= \langle m \rangle $.
We decided to use the dual photon mass to do the 
matching procedure due to the better numerical accuracy of these results  
in \cite{DiGiacomo:1997sm}.
We fix the two parameters $\alpha_0,A$ by requiring that at two points  
of the curve $M_{V}^{latt}(\beta)$ our results coincide with the lattice one.
Since $g_D$ is assumed to correspond to $\sqrt{\beta}$ matching these 
two points determine $\alpha_0$ and $A$.

Let us now summarize the way we proceed

\noindent
1.- The gauge coupling constant is matched to the lattice $g_D ( \mu = 1 )
\equiv \sqrt{\beta}$

\noindent
2.- The renormalization group is used to run upwards to the scale where the
effective potential derived from duality arguments is valid $g_D ( \mu = 1 )
\to R.G.E. \to g_D ( \mu = {D_0} )$ and we require $\lambda ( \mu =
{D_0} ) = 2 g_D^2 ( \mu = {D_0} )$

\noindent
3.- We use the renormalization group to run downwards the scalar coupling
$\lambda ( \mu = {D_0} ) \to R.G.E. \to \lambda ( \mu = 1 )$
At this step we obtained the before mentioned relation between $\lambda$ 
and $g_D$ at the lattice scale.

\noindent
4.- We
fix the values of  $\alpha_0(\mu = {D_0}) $ and $A(\mu=D_0)$ so as to 
match the lattice curve for the dual photon mass at two points $\beta_1,\beta_2$, in the sense $M_v^{latt}(\beta_1)=M_V(\alpha_0,A,\sqrt{\beta_1}); \ \ M_v^{latt}(\beta_2)=M_V(\alpha_0,A,\sqrt{\beta_2})$. 
In our scenario the order parameter for the Coulomb to confined transition is $\langle m\rangle$. 
The appeareance of a non-trivial 
(i.e. away from the origin) minimum will
signal the Higgs phase in the dual (continuum) model and the confined phase on
the lattice.

We have to notice that even if not explicitly written, the field $m$ also runs
down to $\mu = 1$ as $m = m_0 / \sqrt{Z_m ( \mu )}$ . We take $Z_m (
{D_0} ) = 1$.

\begin{figure}[h]\begin{center} 
\epsfig{file=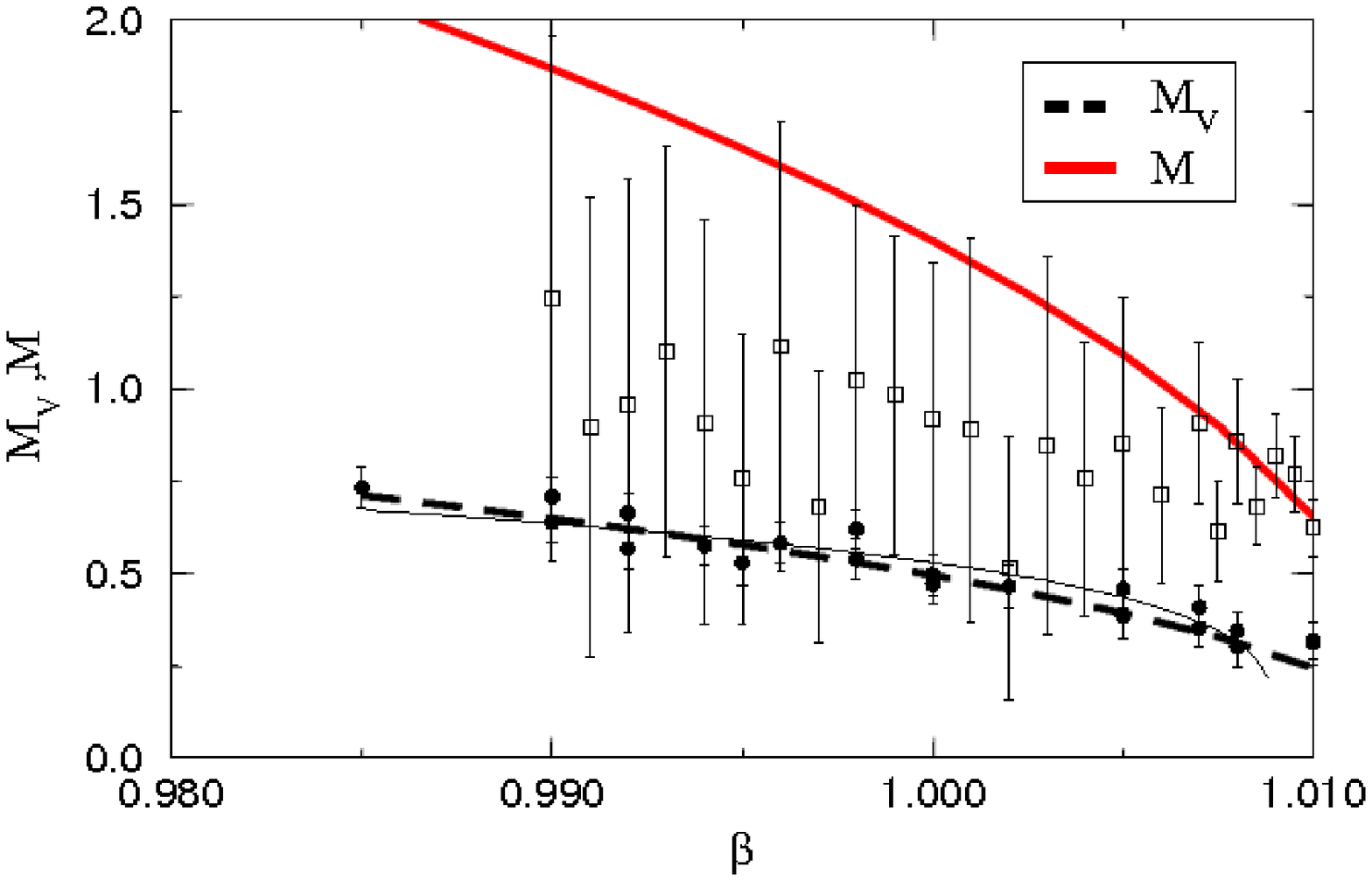,height=10cm}\end{center}
\caption{The figure shows the lattice results for the mass of the 
 dual photon $J^P=1^+$ (dots) and the monopole $J^P=0^+$ (squares) 
versus $\beta$ from \cite{DiGiacomo:1997sm}\label{original} and their 
fit to the former assuming criticality (thin solid line). The thick lines 
correspond to the prediction from our model after the matching procedure
discussed in the text. Two points of the curve $M_V(\beta)$ measured 
on the lattice are used to fix the free parameters in the model we are using. 
This allows to   predict  the critical exponent for $M_V(\beta)$ and 
the curve $M(\beta)$ describing the monopole mass. Clearly our model 
(with a weak first order transition)
describes well the dual photon data and gives a very reasonable 
description of the monopole mass. }
\end{figure}

Doing this we are able to compare our results with the ones obtained in lattice simulations (shown in figure  \ref{original}) and make predictions about the critical exponent 
of the curve of the dual photon  mass as shown in  figure \ref{fotfit}: 
the pseudo-critical exponent emerging from our fit is  
$\nu^{\ast}\simeq 0.46(1)$.

After matching $\alpha_0$ and $A$ everything else is defined and we can
now obtain the curve $M(g_D)$ for the monopole mass as shown in 
fig. \ref{monfit} and fig. \ref{original}. 
A value for the pseudo-critical  exponent $\nu^{\ast}=0.49(1)$ is
obtained. The uncertainty reflects the freedom in choosing $D_0$ and the 
numerical errors of the points we use to fix the parameters; we 
take $D_0$ in the range $10^2 - 10^3$ in lattice units. Larger values
translate in inconsistencies of the one-loop RGE. 
Figure \ref{mon2fas} shows the evolution of the monopole mass across the
phase transition.
\begin{figure}[h]\begin{center}
\epsfig{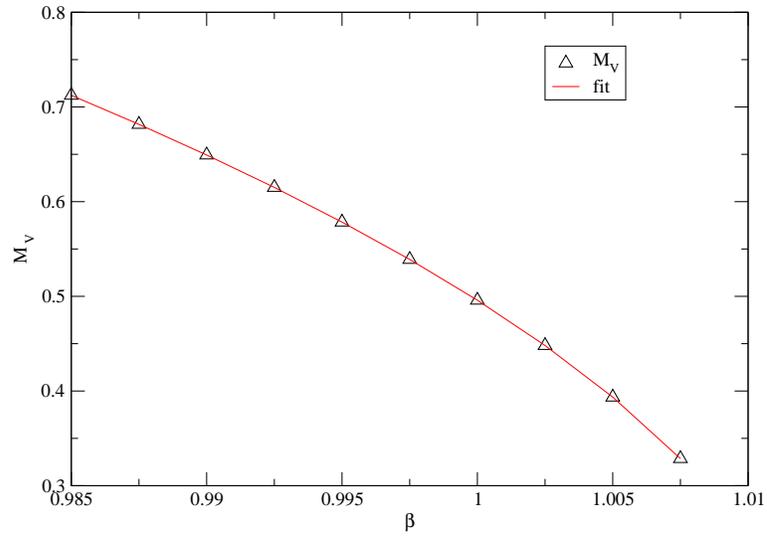}

\caption{Dual photon mass versus $\beta$ and fit. \label{fotfit}}\end{center}
\end{figure}

\begin{figure}[h]\begin{center}
\epsfig{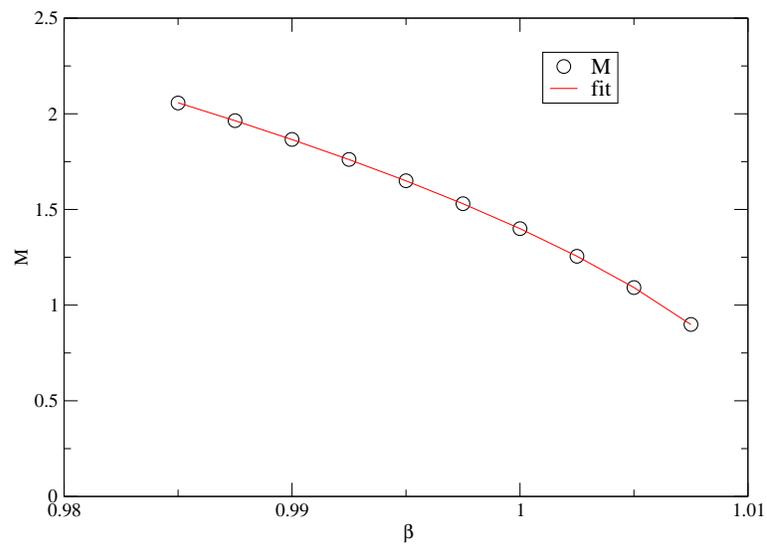}
\caption{Monopole mass obtained from the continuum model 
versus $\beta$ and fit. \label{monfit}}
\end{center}
\end{figure} 
\begin{figure}[h]\begin{center}
\epsfig{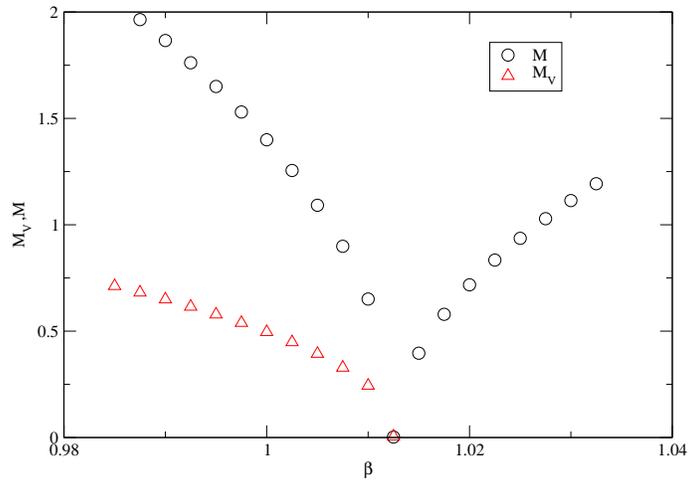}
\caption{The  figure shows the results we obtain for the monopole 
and the dual photon masses across the transition. Even if at the transition the masses seems to vanish they have a non zero value $\mathcal{O}(10^{-2})$ fingerprint of the first order nature of this  transition. This small value compared with the other masses  supports  an effective theory approach of the transition and arises from  its  weak nature (fluctuation  induced) . %The second figure shows the lattice results 
%for the mass of the gaugeballs $J^P=0^+$ (lower line) and $J^P=1^+$ 
%(upper line) 
%versus $\beta$ in both phases from \cite{Cox:1997wd}
\label{mon2fas}}
\end{center}
\end{figure}

The monopole condensate can also be
obtained and it is plotted in fig.\ref{confit}. Extracting 
the pseudo-critical 
exponent we obtain $\delta^{\ast}\simeq 0.48(1)$.  
The results of \cite{DiGiacomo:1997sm} 
lead to  $\delta^{\ast}=1.1 \pm 0.2$. In 
figure \ref{confit}(b) we show the results for the condensate obtained 
using the 
Villain action \cite{Jersak:1999nv}. It is found in these 
simulations that the
critical index is $\delta^{\ast}=0.197(3)$.
Clearly there is a lot of room
for improvement in the numerical simulation field., particularly in what 
concerns the monopole. 
Our prediction on
the other hand is  quite clear; the exponent for the condensate 
should also be Gaussian.

Our results and predictions are also questionable. Apart from the main
conjecture about the form of the quantum corrections due to the $N=1$
breaking, that we regard as a mild one, we have derived our results using
the one-loop effective potential and one-loop beta functions. Some 
conclusions are actually independent of this, for instance the existence of
a first-order phase transition due to the Coleman-Weinberg phenomenon. But
detailed numerics do of course depend on this approximation. The fact 
is that we are always in the weakly coupled phase so the results should
be trustable with a typical 'perturbative' error which indeed seems to be
reasonably small.

\begin{figure}[h]\begin{center}
\epsfig{file=confit.eps,width=10cm}
\epsfig{file=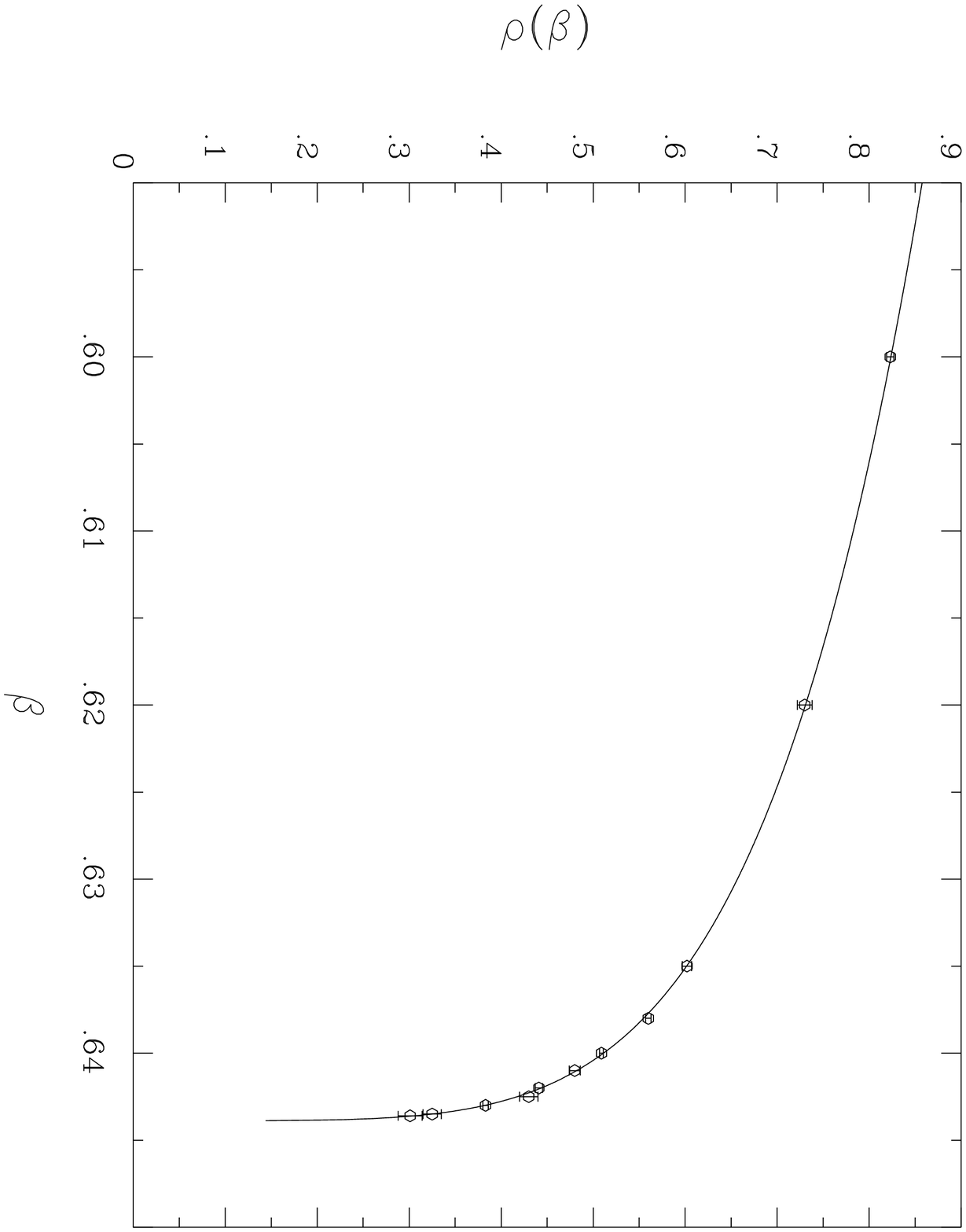,width=8.5cm,angle=90}
\caption{Monopole condensate  versus $\beta$ and fit \label{confit}
and the analogue result obtained in \cite{Jersak:1999nv}. The lattice
results are obtained using the Villain action. }\end{center}
\end{figure}

To conclude,
the arguments based on duality provide a well defined form of the long-distance
effective action, including a well prescribed form of the dependence of
the mass parameted in $g_D$, the relation between the scalar coupling $\lambda$
constant and $g_D$ and so on. This form of the effective action
seems to reproduce well the behaviour observed in lattice simulations.
A clear prediction of our model is that the phase transition is of
first order everywhere.

We have obtained a  good qualitative agreement with the 
(still rather crude) results 
of the lattice simulations concerning mass of the  
dual photon, the monopole mass, 
the monopole condensate and its pseudocritical exponents.

We believe that our results show that the dual Higgs mechanism is at work 
in lattice $U(1)$. 

A very interesting aspect is to understand the relation between  the 
spectrum of the gaugeballs measured in the $U(1)$ lattice gauge theories  
\cite{Cox:1997wd,Majumdar:2003xv}, and the spectrum of the dual abelian 
Higgs model we are considering.
In \cite{Ambjorn:1997cq} a correspondence based on the identification of 
the quantum number  $J^P$ was proposed. While this  argument  for  
the gauge ball state $1^+$ might be correct, we think that it 
cannot safely be extended to the identification 
of the  $0^+$ gaugeball with the monopole since the former carries no magnetic 
charge. The identification is thus questionable. 

On the other end  we would expect that 
a well defined effective theory must include all the light degrees of 
freedom appearing in the original model. So we expect that there should be 
an interpretation of the above gaugeball states in terms of 
monopole-antimonopole  weakly bound states, but 
for the moment we cannot give a precise statement that fulfills our 
expectations.
We think that this problem deserves further study.  

We regard
these explorations as being both interesting and urgent to enlarge our 
understanding of the mechanisms behind confinement.

\section*{Acknowledgements}

We acknowledge the financial support from projects FPA2001-3598,
2001SGR-00065
and EUROGRID (HPRN-CT-1999-00161). 
L.T. whises to acknowledge  A.Dominguez,
Ll.Masanes and A. Prats for useful discussions, 
suggestions and observations on the topic and the hospitality of the theory group of the Physic department at Pisa University where the last version of this work was completed. 
{}
\end{document}